\documentclass[sigconf]{acmart}

\usepackage{geometry}
\usepackage{bm}
\usepackage{algorithm2e}

\usepackage{subcaption}

\AtBeginDocument{%
  \providecommand\BibTeX{{%
    \normalfont B\kern-0.5em{\scshape i\kern-0.25em b}\kern-0.8em\TeX}}}

\acmConference[SIGSPATIAL'22]{The 30th International Conference on Advances in Geographic Information Systems}{November 1-4, 2022}{Seattle, WA, USA}

\setcopyright{none}

\begin{document}


\title[Region2Vec: Community Detection on Spatial Networks Using Graph Embedding]{Region2Vec: Community Detection on Spatial Networks Using Graph Embedding with Node Attributes and Spatial Interactions}


\author{Yunlei Liang}
\affiliation{%
  \institution{Geospatial Data Science Lab, University of Wisconsin}
  \city{Madison, WI}
  \country{USA}}
\email{yunlei.liang@wisc.edu}

\author{Jiawei Zhu}
\affiliation{%
  \institution{School of Geosciences and Info-Physics, Central South University}
  \city{Changsha}
  \country{China}
}
\email{jw\_zhu@csu.edu.cn}

\author{Wen Ye}
\affiliation{%
  \institution{Geospatial Data Science Lab, University of Wisconsin}
  \city{Madison, WI}
  \country{USA}
}
\email{wye35@wisc.edu}

\author{Song Gao}
\affiliation{%
  \institution{Geospatial Data Science Lab, University of Wisconsin}
  \city{Madison, WI}
  \country{USA}
}
\email{song.gao@wisc.edu}

\renewcommand{\shortauthors}{Liang et al.}

\begin{abstract}

Community Detection algorithms are used to detect densely connected components in complex networks and reveal underlying relationships among components. As a special type of networks, spatial networks are usually generated by the connections among geographic regions. Identifying the spatial network communities can help reveal the spatial interaction patterns, understand the hidden regional structures and support regional development decision-making. Given the recent development of Graph Convolutional Networks (GCN) and its powerful performance in identifying multi-scale spatial interactions, we proposed an unsupervised GCN-based community detection method \textit{region2vec} on spatial networks. Our method first generates node embeddings for regions that share common attributes and have intense spatial interactions, and then applies clustering algorithms to detect communities based on their embedding similarity and spatial adjacency. Experimental results show that while existing methods trade off either attribute similarities or spatial interactions for one another,  \textit{region2vec} maintains a great balance between both and performs the best when one wants to maximize both attribute similarities and spatial interactions within communities. 


\end{abstract}





\maketitle

\section{Introduction}

Many real world phenomena happen in the form of networks or can be represented by networks. For example, in a spatial network, nodes are usually geographic locations or regions, and edges are the spatial interactions between different places \cite{gao2013discovering}. The spatial interactions can have various meanings, such as human movements or goods transportation. To extract useful knowledge from such complex networks, community detection algorithms have been widely used. A distinct characteristic of spatial networks is that the nodes may have inherent geographic relationships across different scales. Therefore, two adjacency matrices can be built for spatial networks, the first one represents the flow connections (spatial interactions), and the second one represents the geographic closeness (spatial distribution). 

The convolutional network-based models provide an ideal approach to model the geographic closeness relationship. Graph Convolutional Networks (GCNs) combine both node features and edge relationships through convolution layers, and generate latent features of nodes by aggregating the neighboring relations among nodes \citep{Kipf2017}. However, there are two main issues with applying GCN to community detection. First, existing GCN models are usually supervised or semi-supervised, while community detection is essentially an unsupervised learning problem \citep{Jin2019a}. Second, GCN is not initially designed for community detection and the embedding from GCN is not community-oriented \citep{Jin2019a}. The goal of GCN learning should also be adjusted so that the characteristics of communities can be included.

To solve the abovementioned issues, we proposed a GCN-based unsupervised learning method by designing a community-oriented loss and considering both spatial interaction and geographic characteristics. Especially, we combine information from nodes, edges, neighborhoods, and multi-graphs in the GCN model to effectively learn the graph embedding. Additional clustering is then applied as a post-processing step to discover communities. We call this method \textit{region2vec}. Although this name was first used by \citet{xiang2020region2vec} to detect urban land use type, we would like to extend the concept of \textit{region2vec} and use it to indicate a category of methods that generate latent feature representations based on regions' characteristics. We will demonstrate the effectiveness of our proposed method in community detection tasks on spatial networks.


\section{Methodology}

\subsection{Notations and Problem Definitions}
Graph $\bm{G} = (V, E)$ is defined via a set of nodes $V = (v_1, ..., v_n)$, $|V| = n$ and edges $E$ with $e_{ij} = (v_i, v_j)$. $\bm{A} = [a_{ij}]_{n \times n}$ is an adjacency matrix, where $a_{ij} = 1$ if $e_{ij} \in E$, otherwise $a_{ij} = 0$. $\bm{S} = [s_{ij}]_{n \times n}$ is a spatial interaction matrix, where $s_{ij}$ represents the flow intensity between nodes $v_i$ and $v_j$. An $ n \times m$ attribute matrix $\bm{X}$ is used to denote the multidimensional attributes of nodes.

The community detection aims to partition the $n$ nodes into $K$ communities $\{C_1, C_2, ..., C_K\}$ and each node will have a label $c_i$ indicating its community membership, $c_i \in \{1, 2, ..., K\}$.

\subsection{Data}

The major data source used for building the spatial network (graph) is the SafeGraph business venue database\footnote{\url{https://www.safegraph.com}}. SafeGraph collects over 8 million points of interest (POIs) with visit patterns in the U.S. 
To construct the spatial flow network in this study, all the place visits are aggregated to the census tract level \citep{kang2020multiscale}. The census tracts are then used as the nodes, and the human movement flows are the edges with flow intensity as the weights. 

The spatial adjacency matrix is built based on the geographic relationship among census tract boundaries from the TIGER/Line Shapefiles\footnote{\url{https://www.census.gov/geographies/mapping-files/time-series/geo/tiger-line-file.html}}. We specifically use the Rook-type contiguity relationship, which defines neighbors by the existence of sharing edges. Only census tracts with shared borders larger than zero meters will be considered as spatially adjacent. The node attributes are collected from U.S. Census American Community Survey (ACS) 2015-2019 5-year estimates. Features including poverty population, race/ethnicity, and household income are used in the model.

\begin{figure*}[h]
	\includegraphics[width=0.76\linewidth]{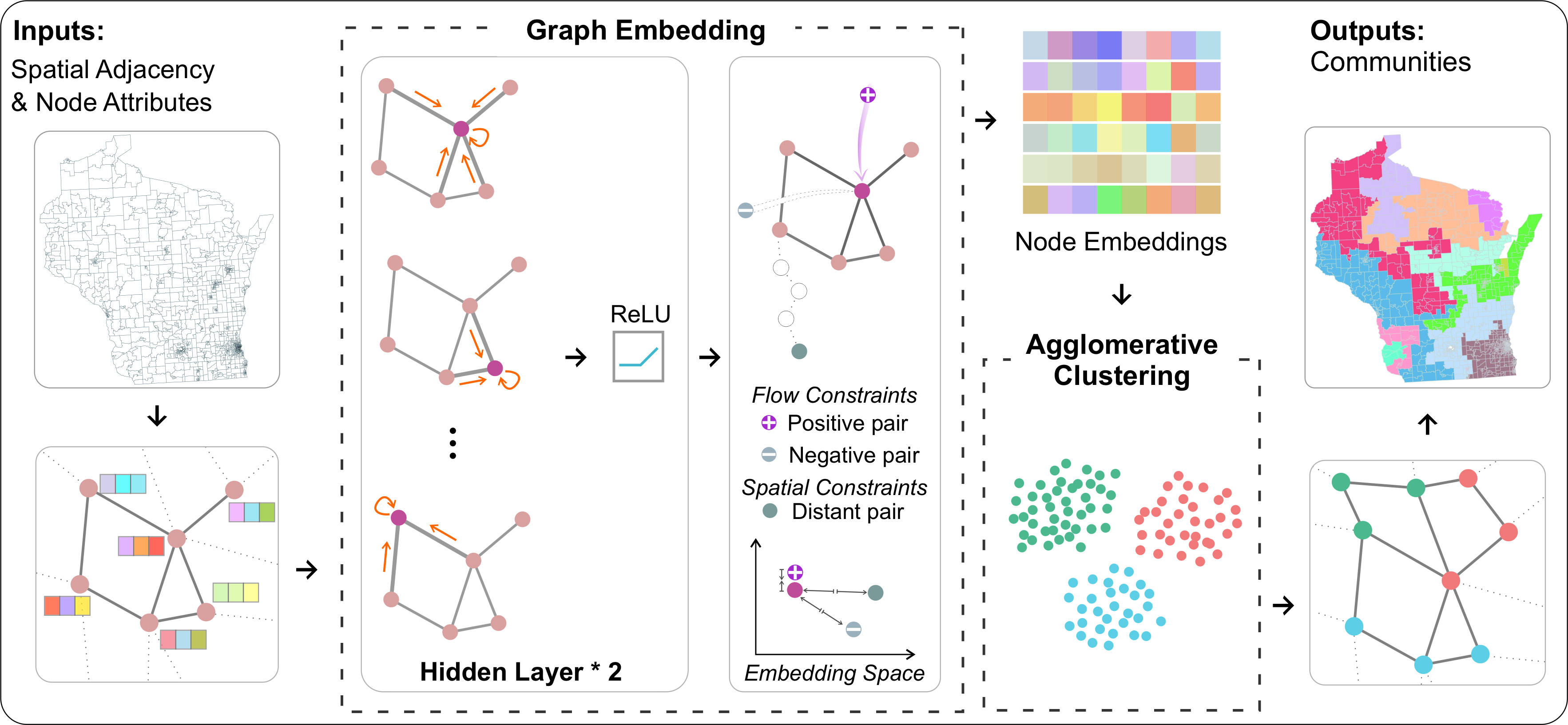}
	\caption{The workflow for community detection using the regions2vec method}
	\label{fig:workflow}
\end{figure*}

\subsection{Algorithm}

Our goal is to identify spatial network communities, where nodes (i.e., geographic regions) within the same community satisfy the following expectations: 1) share similar attributes; 2) have intense spatial interactions; and 3) are spatially contiguous. To achieve this, we proposed a two-stage community detection algorithm using both node attributes and spatial interactions. The workflow is shown in Figure \ref{fig:workflow}.

\subsubsection{Stage One: Node Representation Learning}

One special characteristic of spatial networks is that the nodes that are spatially adjacent tend to be similar in attributes, according to the first-law of geography (spatial dependency effect) \citep{tobler1970computer}. Since the critical operation of graph convolutional neural networks is to aggregate neighbor information, making it a natural tool that fits this characteristic when given spatial adjacency matrix and node attributes as inputs.

As we define $Z^{(1)}$ and $Z^{(2)}$ as the outputs of the first and second graph convolutional layers, and $W^{(0)}\in \mathbb{R}^{m\times n_{hidden}}$ and $W^{(1)}\in \mathbb{R}^{n_{hidden}\times n_{output}}$ as the weights of two layers, the forward propagation model can be formalized as Equation \ref{eq:gc}:
\begin{equation}
\begin{aligned}
Z^{(1)}=ReLU (\tilde{D}^{-\frac{1}{2}}\tilde{A}\tilde{D}^{-\frac{1}{2}}X{{W}_{0}});
Z^{(2)}= \tilde{D}^{-\frac{1}{2}}\tilde{A}\tilde{D}^{-\frac{1}{2}}Z^{(1)}{{W}_{1}}
\end{aligned}
\label{eq:gc}
\end{equation}
where $A$ and $I$ are the spatial adjacency and identity matrices, $\tilde{A}=A+I$, and $\tilde{D}$ is the degree matrix of $\tilde{A}$. 

However, GCN is a semi-supervised model and not community-oriented. So we utilize the spatial interaction flow strength and geographic distance as constraints to guide the learning process. Specifically, the nodes without flow interactions are considered as negative pairs and are pushed away in the embedding space, while those with interactions will be treated as positive pairs and drawn closer; the greater the flow intensity is, the closer we bring them together in the embedding space. Moreover, we set a threshold to push away node pairs that are spatially distant from each other to guarantee the spatial contiguity. Thus, the loss function is designed as Equation \ref{eq:loss}:
\begin{equation}
\begin{aligned}
L_{hops} = \sum{\frac{\mathbb{I}(hop_{ij} > \epsilon)d_{ij}}{\log(hop_{ij})}};
Loss= \frac{\sum_{p=1}^{N_{pos}}{\log(s_{p})d_{pos_{p}}}/N_{pos}}{\sum_{q=1}^{N_{neg}}{d_{neg_q}}/N_{neg} + L_{hops}},
\label{eq:loss}
\end{aligned}
\end{equation}
where $hop_{ij}$ represents the hop numbers of the shortest path between $v_i$ and $v_j$ in the graph, and $d_{ij}$ is the euclidean distance between the corresponding embedding representations. $\mathbb{I}(\cdot)$ is set to 1 if $hop_{ij} > \epsilon$, or 0 otherwise. Positive pairs and negative pairs of nodes are denoted by $pos_p, p\in [0,N_{pos}]$ and $pos_q, q\in [0,N_{neg}]$, respectively. Since the intensity of flow $s_p$ has a large range of values, we adopt a log transformation so that the flow values will not get overwhelmed by the extremely large values. The pseudo code of \textit{region2vec} is shown in Algorithm \ref{alg:one}.

\RestyleAlgo{ruled}
\setlength{\textfloatsep}{0pt}%
\begin{algorithm}[ht]
	\caption{Region2Vec}\label{alg:one}
	\SetKwInput{KwInput}{Input}
	\SetKwInput{KwOutput}{Output}
	\KwInput{ $\bm{G}$; $\bm{A}$; $\bm{S}$; $\bm{X}$; $hops_{i,j}, \forall i,j\in V$ and threshold $\epsilon$; number of layers $L$; weight matrices $W^l,\forall l \in \{0,\cdots,L-1\}$}
	\KwOutput{Node representations $\bm{z}_v$ for all $v\in V$}
	$Z^{(0)} \gets \bm{X}$\;
	$\tilde{A} \gets A+I$\;
	$pos_m \gets (i,j)$, for all $s_{ij} > 0$\;
	$neg_n \gets (i,k)$, for all $s_{ik} = 0$\;
	\For{each iteration}{
		\For{$l=0,\cdots,L-1$}{
			$Z^{(l+1)}=ReLU (\tilde{D}^{-\frac{1}{2}}\tilde{A}\tilde{D}^{-\frac{1}{2}}Z^{(l)}{{W}^{l}})$\;
			
		}
		$d_{ij} = \|z_i-z_j\|$\;
		$L_{hops} = \sum{\mathbb{I}(hop_{ij} > \epsilon)d_{ij}/\log(hop_{ij})}$\;
		$Loss = \frac{1}{N_{pos}}\sum_{p=1}^{N_{pos}}{\log(s_{p})d_{pos_{p}}}/
		(\frac{1}{N_{neg}}\sum_{q=1}^{N_{neg}}{d_{pos_{q}}}+L_{hops})
		$\;
		Compute $g \gets \nabla Loss$\;
		Conduct Adam update using gradient estimator $g$
	}
	$\bm{z}_v \gets z_v^{L}, \forall v\in V$
\end{algorithm}

\subsubsection{Stage two: The Agglomerative Clustering}

After obtaining the node representation, agglomerative clustering is utilized. Agglomerative clustering uses a bottom-up approach: each node is treated as a separate cluster at the beginning, and then is merged successively into groups \citep{scikit-learn}. The merge criterion is measured using the linkage type ``Ward", which minimizes the sum of squared differences within all clusters and can generate clusters with the most regular sizes compared with other types \citep{scikit-learn}.

Another advantage of using agglomerative clustering is that it supports the incorporation of connectivity constraints \citep{scikit-learn}. This characteristic is especially critical in our study as the spatial contiguity is an inherent requirement of community detection in spatial networks. The spatial adjacency matrix is used to preserve the spatial contiguity and impose local structures.

\subsection{Baseline Algorithms}
We use a variety of baseline algorithms to compare their performances with our proposed method \textit{region2vec}. 

\par{\textit{Louvain}:
	The Louvain algorithm \citep{blondel2008fast} is a heuristic method based on modularity optimization. It is applied to identify communities only using flow connections.}

\par{\textit{Random walk based models}:
	Two random walk based graph embedding models, Deepwalk \citep{perozzi2014deepwalk} and Node2vec \citep{grover2016node2vec} are used to learn continuous feature representations. The two methods are conducted using the spatial adjacency matrix as the input, followed by the same agglomerative clustering algorithm.}

\par{\textit{LINE}:
	Large-scale Information Network Embedding (LINE) \citep{tang2015line} is a network embedding method suitable for arbitrary types of information networks especially with large sizes. LINE uses the spatial adjacency matrix as the input and is followed by agglomerative clustering.}

\par{\textit{K-Means}:
	The K-Means clustering algorithm aims to group nodes based on their feature similarities \citep{macqueen1967some}. The K-Means clustering algorithm is directly applied on the node multidimensional attributes but it does not consider the graph structure. 
}

\subsection{Evaluation Metrics}
To compare the performance of all the methods, the following metrics are used to comprehensively evaluate the communities. 
\par{\textit{Intra/Inter Flow Ratio}:
	The spatial interaction flow ratio is specifically designed for this study, it measures the ratio of edge weights sum within each community (intra-flow weights) when ${c_i=c_j}$ and the edge weights sum between different communities (inter-flow weights) when ${c_i \neq c_j}$, which is similar to the concept of modularity \citep{newman2006modularity}. As shown in Equation \ref{eq:flow_ratio}, $s_{ij}$ represents the flow intensity between two nodes $i$ and $j$. 
	\begin{equation}
	\begin{gathered}
	Intra/Inter \ Flow \ Ratio = \frac{\sum_{c_i=c_j} s_{ij}}{\sum_{c_i \neq c_j} s_{ij}}; 
	c_i, c_j\in {1, 2,\cdots, K} 
	\end{gathered}
	\label{eq:flow_ratio}
	\end{equation}
}

\par{\textit{Inequality}:
	The inequality metric was proposed by \citet{pandey2022infrastructure} to measure the infrastructure inequality across multiple geographic regions (Equation \ref{eq:inequality}). $\sigma$ is the standard deviation and $\mu$ is the mean. A value of 1 indicates maximum inequality, and 0 indicates no inequality. 
	\begin{equation}
	I = \frac{\sigma}{\sqrt{\mu (1-\mu)}}; 0 < \mu < 1.
	\label{eq:inequality}
	\end{equation}
}

\par{\textit{Similarity Metrics}:
	The cosine similarity is used to calculate the L2-normalized dot product of vectors \citep{scikit-learn}.
}

\par{\textit{Homogeneity Scores}:
	The homogeneity score is used to evaluate if nodes in each community are more homogeneous and have more similar socio-economic characteristics. It is calculated based on the percentage of the population with income at or lower than 200\% federal poverty level. 
}

\section{Results}


\begin{table}
	\centering
	\caption{The metrics comparison of all methods. (In bold: best; \underline{Underline}: second best)}
	\label{tab:metrics}
	\resizebox{\linewidth}{!}{
		\begin{tabular}{ccccc} 
			\hline
			Methods    & \begin{tabular}[c]{@{}c@{}}Intra/Inter\\Flow Ratio\end{tabular} & Inequality & \begin{tabular}[c]{@{}c@{}}Cosine\\Similarity\end{tabular} & Homogeneity  \\ 
			\hline
			DeepWalk   & 2.585                                                           & 0.375      & 0.960                                                      & 0.103        \\
			K-Means    & 0.438                                                           & \textbf{0.213}      & \textbf{0.983}                                                     & \textbf{0.515}      \\
			LINE       & 0.273                                                           & 0.723      & 0.872                                                      & 0.012        \\
			Louvain    & \textbf{4.864}                                                           & 0.373      & 0.964                                                      & 0.080        \\
			Node2vec   & 2.717                                                           & 0.437      & 0.951                                                      & 0.091        \\
			Region2vec & \underline{3.588} & \underline{0.367}      & \underline{0.974} & \underline{0.105}        \\
			\hline
	\end{tabular}}
\end{table}


The performance of community detection on spatial networks are compared together for all the introduced methods. The results are based on the number of community of 14, which is the optimal community number in the Louvain algorithm. 


In total four metrics are used to evaluate the performance of these methods on community detection from different perspectives and the results are listed in Table \ref{tab:metrics}. Overall, our proposed \textit{region2vec} method maintains a great balance between attribute similarity and spatial interaction intensity and performs the best when one wants to maximize both attribute similarities and spatial interactions simultaneously within communities.

First, the intra/inter flow ratio represents the ratio of intra-community flows and inter-community flows. The Louvain method, which takes the spatial interaction flow matrix as the only input, has the highest flow ratio value. Our proposed \textit{region2vec} method performs the second best in the spatial interaction perspective. Following them are the two random walk based algorithms: Node2vec and Deepwalk, which have similar ratios. Lastly the K-Means method and the LINE have the lowest ratios.

For the inequality, a lower value represents that the nodes within communities are more similar and have lower variations. We use the median inequality to represent each method in Table \ref{tab:metrics}. The K-Means clustering method has the lowest median inequality as it clusters nodes purely based on their attribute similarity; the nodes in the same cluster have more similar features and therefore, are more equal. The proposed \textit{region2vec} method has the second lowest inequality, meaning that it also has a good performance for grouping nodes with similar features. The remaining four methods have higher inequality as they do not consider feature information.


For the cosine similarity, as expected, K-Means clustering that uses only node attributes in the process performs the best and it has the highest cosine similarity according to Table \ref{tab:metrics}. The \textit{region2vec} method, again, is rated the second best.


Last but not least, as the homogeneity score is a metric evaluating how homogeneous the clusters are in terms of lower-income population percentage, K-Means clustering has the highest score. The \textit{region2vec} has the second highest score, meaning that it is able to better group homogeneous nodes than the other four baselines that are purely based on edge information in graphs.

\section{Conclusions}
This study proposed an unsupervised community detection method called \textit{region2vec} on spatial networks. Using a GCN-based model, \textit{region2vec} considers the spatial adjacency, spatial interaction flows, and the node attributes. Through a community-oriented loss function, this method first generates embedding for nodes based on attribute similarity and flow interactions. Communities are further identified through agglomerative clustering with the spatial adjacency constraint. The \textit{region2vec} method has been compared with the most commonly used community detection methods and shown a great performance when considering both node attributes and spatial interactions. Our future work will apply the proposed method to the regionalization problems such as the rational service area development in public health and the redistricting problem in political science. 

This research demonstrates the good potential of graph embedding and GCN in the community detection on spatial networks, as well as the integration of geospatial constraints in deep learning models, which can contribute to the increasing interests on GeoAI development in the SIGSPATIAL community \citep{hu2019geoai,janowicz2020geoai}.


\begin{acks}
We acknowledge the funding support from the County Health Rankings and Roadmaps program of the University of Wisconsin Population Health Institute, Wisconsin Department of Health Services, and the National Science Foundation funded AI institute [Grant No. 2112606] for Intelligent Cyberinfrastructure with Computational Learning in the Environment (ICICLE). Any opinions, findings, and conclusions or recommendations expressed in this material are those of the author(s) and do not necessarily reflect the views of the funders.
\end{acks}

\bibliographystyle{ACM-Reference-Format}
\bibliography{references.bib}

\end{document}